\documentclass [12pt]{article}
\usepackage{graphicx,amssymb,amsfonts,latexsym,amsmath,amsthm,times}
\usepackage{epsfig}
\usepackage{fancyhdr}
\usepackage{color}
\usepackage[english]{babel}
\numberwithin{equation}{section}

\begin{document}

\footnotesize {\flushleft \mbox{\bf \textit{Math. Model. Nat.
Phenom.}}}
 \\
\mbox{\textit{{\bf Vol. 6, No. 1, 2011, pp. 138-148 }}}

\medskip

\noindent DOI: { 10.1051/mmnp/20116107}

\thispagestyle{plain}

\vspace*{2cm} \normalsize \centerline{\Large \bf Pattern Formation
Induced by Time-Dependent Advection}

\vspace*{1cm}

\centerline{\bf A. V. Straube$^{1,2}$\footnote{Corresponding
author. E-mail: straube@physik.hu-berlin.de} and A. Pikovsky$^2$ }

\vspace*{0.5cm}

\centerline{$^1$ Department of Physics, Humboldt University of Berlin,}
 \centerline{Newtonstr. 15, D-12489, Berlin, Germany}

\centerline{$^2$ Department of Physics and Astronomy, University of Potsdam,} \centerline{Karl-Liebknecht-Str. 24/25, D-14476 Potsdam-Golm,
Germany}


\vspace*{1cm}

\noindent {\bf Abstract.} We study pattern-forming instabilities in reaction-advection-diffusion systems. We develop an approach based on Lyapunov-Bloch exponents to figure out the impact of a spatially periodic mixing flow on the stability of a spatially homogeneous state. We deal with the flows periodic in space that may have arbitrary time dependence. We propose a discrete in time model, where reaction, advection, and diffusion act as successive operators, and show that a mixing advection can lead to a pattern-forming instability in a two-component system where only one of the species is advected. Physically, this can be explained as crossing a threshold of Turing instability due to effective increase of one of the diffusion constants.

\vspace*{0.5cm}




\vspace*{1cm}

\setcounter{equation}{0}
\section{Introduction}

Reaction-diffusion systems is a well-established class of models describing various aspects of pattern formation far from equilibrium. Quite often pattern-forming fields are transported by fluid flows -- examples range from a development of plankton patterns in oceanic flow \cite{huisman-etal-06} to chemical reactions in microchannels \cite{leconte-etal-04}. By incorporating the flow in the model, one arrives at reaction-advection-diffusion models, with a much richer variety of possible phenomena (see, e.g., a recent review paper~\cite{Tel_etal-05}). In Ref.~\cite{Rovinsky-Menzinger-94} it was demonstrated that a differential flow (where only some components are advected) may lead to new instabilities in the system, in particular to new convective instabilities~\cite{Yakhnin-Rovinsky-Menzinger-95}. This study was restricted to a planar geometry, but later in \cite{Khazan-Pismen-95, Balinsky-Pismen-98} such an instability was demonstrated for a circular geometry (Couette flow) as well. As has been recently shown, a simple \textit{shear} flow is able to destabilize the spatially homogeneous state \cite{vasquez-04}. A reasonable question to ask is whether a similar destabilization effect can be found for \textit{mixing} flows. One might intuitively expect that as mixing smears spatial nonuniformities, it results in stabilization of a spatially homogeneous state (like it happens if the reaction is chaotic~\cite{Straube-Abel-Pikovsky-04}). In this paper we particularly address this question by presenting an approach to study pattern-forming instabilities in periodic in space mixing flows. It is based on the calculation of Lyapunov-Bloch exponents (cf.~\cite{pikovsky-89}), and provides an efficient tool for finding mostly unstable patterns. As a particular example we consider the effect of a mixing advection on a general two-dimensional reaction-diffusion system capable of Turing instability and demonstrate that  instability can be induced by an advection of one component of the reaction.

\setcounter{equation}{0}
\section{Model formulation}
\subsection{Continuous-time model}

A variety of situations in biological and chemical contexts can be described by the dynamics of two interacting species -- an activator and an inhibitor. A popular model, e.g., is the Brusselator~\cite{nicolis-prigogine-77} or its modifications. For spatially distributed fields, the dynamics also includes diffusion terms (molecular diffusion for chemical systems or irregular mobility in biological applications) and advection due to an imposed macroscopic velocity field ${\textbf V}({\textbf r},t)$. We assume the latter to be incompressible, furthermore we normalize time by the characteristic advection time. In this paper we are interested in stability properties of a steady homogeneous distribution. Denoting the deviations of the concentrations of the species from this steady state with $P$, $Q$, we  arrive at a linear reaction-advection-diffusion system governed by a couple of dimensionless equations
\begin{eqnarray}
\frac{\partial P}{\partial t}+ {\textbf V}\cdot \nabla P & =
& D_P \nabla^2 P+a P+b Q, \label{eq_X} \\
\frac{\partial Q}{\partial t}+{\textbf V}\cdot \nabla Q & = & D_Q
\nabla^2 Q+cP+dQ. \label{eq_Y}
\end{eqnarray}
\noindent Here $P$ and $Q$ are deviations of chemical concentrations from the steady state, $D_P$ and $D_Q$ are their
corresponding diffusivities; $a,b,c,d$ are parameters of the kinetics. We assume that concentrations do not influence the flow, so that the velocity field ${\textbf V}({\textbf r},t)$ does not depend on $P$ and $Q$.

In the absence of advection the problem reduces to the classical
reaction-diffusion model (see, e.g., Ref.~\cite{pismen-book}),
where two principal instabilities are a spatially homogeneous Hopf bifurcation and a Turing instability~\cite{turing-52}. In this paper our main interest is in the case where Turing instability is dominant, what requires in particular
that the diffusion constants $D_P$ and $D_Q$ are different. Our goal is to describe unstable modes in the presence of advection term in
(\ref{eq_X}), (\ref{eq_Y}). This can be done numerically, and for an effective calculation we formulate a discrete in time model of the reaction-advection-diffusion system above.

%
\subsection{Discrete in time model}

\begin{figure}[!h]
\centering
\includegraphics[width=12.0cm]{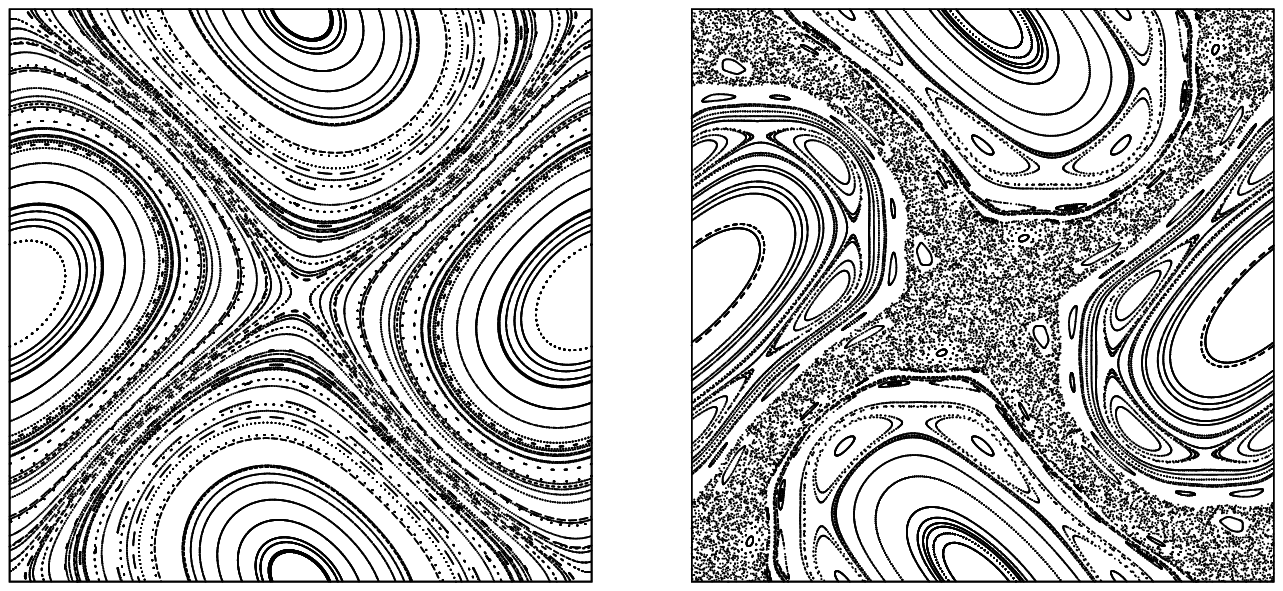}
\caption{Phase portraits of the map (\ref{advmap-x}),
(\ref{advmap-y}) for periodic driving, $T_1=T_2=0.5$,
$\theta_x=\theta_y=0$. The parameters $U_x=U_y=1.5$ (left) and
$U_x=U_y=3.0$ (right).} \label{fig:phaseportr}
\end{figure}

A typical experimental realization of a two-dimensional mixing flow ${\textbf V}({\textbf r},t)$ is a flow periodic in space and time,
as realized e.g. in experiments~\cite{Rothstein-Henry-Gollub-99}. Qualitative behavior of the system can be
understood from a simple problem, where advection is modeled by a two-dimensional time-dependent flow which is $2\pi$-periodic in space
and $T$-periodic in time and first introduced by Antonsen et al.~\cite{antonsen-etal-96}
\begin{equation}
  \label{flow}
  {\textbf V} =  {\textbf e}_xU_x f(t) \sin [y+\theta_x(t)]+  {\textbf e}_yU_y [1-f(t)] \sin
  [x+\theta_y(t)],
\end{equation}
The function $f(t)$ describes switching between two shear flows in $x$ and $y$ directions within
time intervals $T_1$ and $T_2$ with the amplitudes $U_x$, $U_y$, respectively
\begin{equation} \label{f_function}
f(t) = \begin{cases} 1, & 0 > t > T_1, \\ 0, & T_1 < t < T_2.
\end{cases}
\end{equation}
The advantage of such a setup is that the trajectories of particles in the flow (\ref{flow}), (\ref{f_function}) can be obtained explicitly. The transformation of particle coordinates due to the advection during one time interval $T=T_1+T_2$ is governed by a map
\begin{eqnarray} \label{advmap-xy}
\bar{x} & = & x+U_x T_1 \sin (y+\theta_x) \;,
\label{advmap-x} \\
\bar{y} & = & y+U_y T_2 \sin (\bar{x}+\theta_y) \;,
\label{advmap-y}
\end{eqnarray}
\noindent where $x=x(t_n)$, $y=y(t_n)$, $\bar{x}=x(t_n+T)$, $\bar{y}=y(t_n+T)$, and $n$ is the number of iteration. The phase
space dynamics of the area-preserving map (\ref{advmap-x}), (\ref{advmap-y}), restricted to the basic domain of periodicity $0\leq x,y\leq 2\pi$ is demonstrated in Fig.~\ref{fig:phaseportr}. It is typical of Hamiltonian flows with regular islands and
domains of chaotic behavior. With growth of advection rate ($U_x$, $U_y$), domains of the quasiperiodic dynamics are gradually
superseded by the regions of the chaotic dynamics.


While considering of the reaction-advection-diffusion system (\ref{eq_X}), (\ref{eq_Y}) with the flow (\ref{flow}),
(\ref{f_function}) remains a complex computational problem, there is a possibility to simplify the analysis,
reducing the system to a discrete-in-time model. In the vein of the approach used in Ref. \cite{pikovsky-03} we consider a model
situation where the action of advection, diffusion, and reaction is separated in time. We assume, that within each time interval
$T$ both scalar fields $P$ and $Q$ evolve in three stages, corresponding to advection, diffusion, and  reaction.

During the first stage, both scalars evolve independent of each
other according to pure advection. For a scalar density, this process can be described in terms of the
Frobenius-Perron operator $L_A$ for the map (\ref{advmap-x}), (\ref{advmap-y}). Then, on the next step, the
diffusion operator $L_{D_{i}} = \exp(T D_i \nabla^2)$ is applied. Finally, the fields are subject to
reaction according to the system
\begin{eqnarray}
\frac{\partial P}{\partial t} & =
& a P+b Q, \label{eqt_X} \\
\frac{\partial Q}{\partial t} & = & cP+dQ. \label{eqt_Y}
\end{eqnarray}
Evolution during time $T$ of this system leads to a linear transformation
\begin{equation}
\begin{pmatrix}\bar P\\ \bar Q\end{pmatrix}=
L_R \begin{pmatrix} P\\  Q\end{pmatrix}=
\begin{pmatrix} \frac{(\lambda_- -a)e^{\lambda_+ T}-(\lambda_+-a)e^{\lambda_- T}}{\lambda_--\lambda_+} & \frac{b(e^{\lambda_- T}-e^{\lambda_+ T})}{\lambda_--\lambda_+}\\
\frac{(\lambda_--a)(\lambda_+-a)(e^{\lambda_+ T}-e^{\lambda_- T})}{b(\lambda_--\lambda_+)} &
\frac{(a-\lambda_+)e^{\lambda_+ T}+(\lambda_--a)e^{\lambda_- T}}{\lambda_--\lambda_+}
\end{pmatrix}
\begin{pmatrix} P\\  Q\end{pmatrix}
\label{oper_A}
\end{equation}
where the exponents
\[
\lambda_{\pm}=\frac{a+d}{2}\pm\frac{\sqrt{(a-d)^2+4bc}}{2}
\]
are assumed to be real, in accordance with our choice of absence of Hopf bifurcation.
The reaction-advection-diffusion propagator over one time
interval is given by the product $L_R L_{D_i} L_A$ of the operators.
The goal of the stability analysis is to find unstable eigenvalues and eigenmodes of this operator.

\setcounter{equation}{0}
\section{Bloch ansatz}

Although the basic flow ${\textbf V}({\textbf r},t)$ is periodic in space, a perturbation of the field has not to be periodic. A general perturbation should  be taken in the Bloch form
\begin{equation} \label{Bloch-ansatz}
P(x,y,t_n)=e^{i\kappa_x x+i\kappa_y y}\Phi(x,y,t_n), \quad
Q(x,y,t_n)=e^{i\kappa_x x+i\kappa_y y}\Psi(x,y,t_n),
\end{equation}
\noindent where the functions $\Phi$, $\Psi$ are $2\pi$-periodic in space and additional parameters $\kappa_x$,
$\kappa_y$ stand for quasimomenta. Since the exponential factor $e^{i\kappa_x x+i\kappa_y y}$ has a period of unit with
respect to $\kappa_x$, $\kappa_y$, we consider a symmetric interval of independent values $\kappa_x, \,\kappa_y \in
[-\frac{1}{2},\frac{1}{2}]$. Then, because of periodicity of functions $\Phi$, $\Psi$, the solutions can be represented as Fourier series
\begin{equation} \label{Fourier-decomp}
\Phi(x,y,t_n)=\sum_{l,m}\phi_{lm}(t_n)\, e^{i(lx+my)}, \quad
\Psi(x,y,t_n)=\sum_{l,m}\psi_{lm}(t_n)\, e^{i(lx+my)}.
\end{equation}
Of all operators $L_A$, $L_D$, and $L_R$ the most nontrivial is the advection operator $L_A$. We derive its Fourier representation in appendix~\ref{appendix2}. The resulting transformation of the Fourier components during the advection and diffusion stages reads
\begin{eqnarray}
L_{D_P}L_A{\phi}_{lm} & = & e^{-[(l+\kappa_x)^2+(m+\kappa_y)^2]D_P
T} \sum_{p,q} g_{lmpq} \, \phi_{pq}, \label{m1}\\
L_{D_Q}L_A{\psi}_{lm}^{AD} & = & e^{-[(l+\kappa_x)^2+(m+\kappa_y)^2]D_Q
T} \sum_{p,q}g_{lmpq} \, \psi_{pq}, \label{m2}\\
g_{lmpq} & = & e^{-i(q-m)\theta_x}e^{-i(p-l)\theta_y} J_{q-m}[U_x
T_1(p+\kappa_x)]J_{p-l}[U_y T_2(m+\kappa_y)], \nonumber
\end{eqnarray}
\noindent where $J_m(z)$ is the Bessel function of the first kind.
These components further interact according to the reaction stage (\ref{oper_A})
(where one should replace $P\to\phi$, $Q\to\psi$). The resulting model is a composition of (\ref{m1}), (\ref{m2}), and (\ref{oper_A}).

\setcounter{equation}{0}
\section{Results of the stability analysis}

In this section we apply the proposed model to study the
influence of the advection on pattern formation in a reaction-advection-diffusion system. To
some extent, the influence of advection in an advection-diffusion
system can be understood from the idea of effective diffusion: mixing
effectively increases the diffusion constant. Therefore one can expect that the dynamics of a reaction-advection-diffusion system is similar to that of an reaction-diffusion system with larger diffusion constants.
The mostly interesting point is that in the system under consideration there are two
coupled species, and the Turing instability is caused by a
difference in diffusivities of species. Although advection can
effectively change the diffusion constants of the species, it is
not clear how this difference will be influenced by advection.
However, the situation becomes much more transparent if only one specie is
advected. Then an advection, contributing to its effective diffusion, may increase or decrease the difference of diffusion constants, thus enhancing or suppressing the instability.

Below we focus on a situation when only one species, namely that of higher diffusion constant, is advected. We set the
parameters $b=8$, $d=-9$, $D_P=0.0025$, $D_Q=0.0075$. So, we assume that the mobility of the ``activator'' $P$ is relatively low,
and it is not advected at all.

To perform a linear stability analysis of spatially homogeneous states with respect  to inhomogeneous perturbations, we apply the method of (transversal) Lyapunov exponents (LE). We use the usual method for estimation of the largest LE of mappings (see, e.g., Ref.~\cite{pikovsky-03}). We start with an arbitrary initial distributions for $\phi$, $\psi$, with vanishing spatial
average, and iterate the mapping, performing a renormalization of the linear fields. The averaging of the logarithm of the normalization factors yields the LE. Note that this method can be equally well applied to both regular and irregular flows (\ref{flow}), (\ref{f_function}). Here, however, we focus on the simplest case of time-periodic advection, when $T_1=T_2=0.5$, $T=1$, $\theta_x=0$, $\theta_y=0$, and also put $U_x=U_y \equiv U$.

\begin{figure}[!h]
\centering
\includegraphics[width=8.0cm]{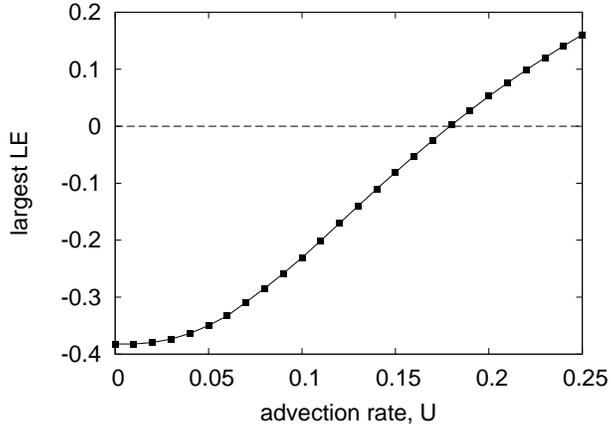}
\caption{Largest Lyapunov exponent as a function of advection rate
$U$. The parameters are $a=5.35$, $c=-6.35$, $\kappa_x=0.5$,
$\kappa_y=0.5$.} \label{fig:LE_adv}
\end{figure}

We choose the parameters of the reaction in such a way that the homogeneous solution is stable in the absence of advection, and then switch on mixing of the specie $Q$. The dependence of the largest LE on the advection rate is presented in Fig.~\ref{fig:LE_adv}.
%
%
%
%
One can see that the impact of advection results in the growth of the largest LE, which  becomes positive at $U_{cr}\approx0.18$.
So, this example clearly demonstrates that mixing can play a destabilizing role.

Remarkably, the quasimomenta in the Bloch ansatz~(\ref{Bloch-ansatz}) are essential in the stability analysis. Below we present three examples where mostly unstable modes correspond to different values of quasimomenta.

\begin{figure}[!h]
\centering
\includegraphics[width=12.0cm]{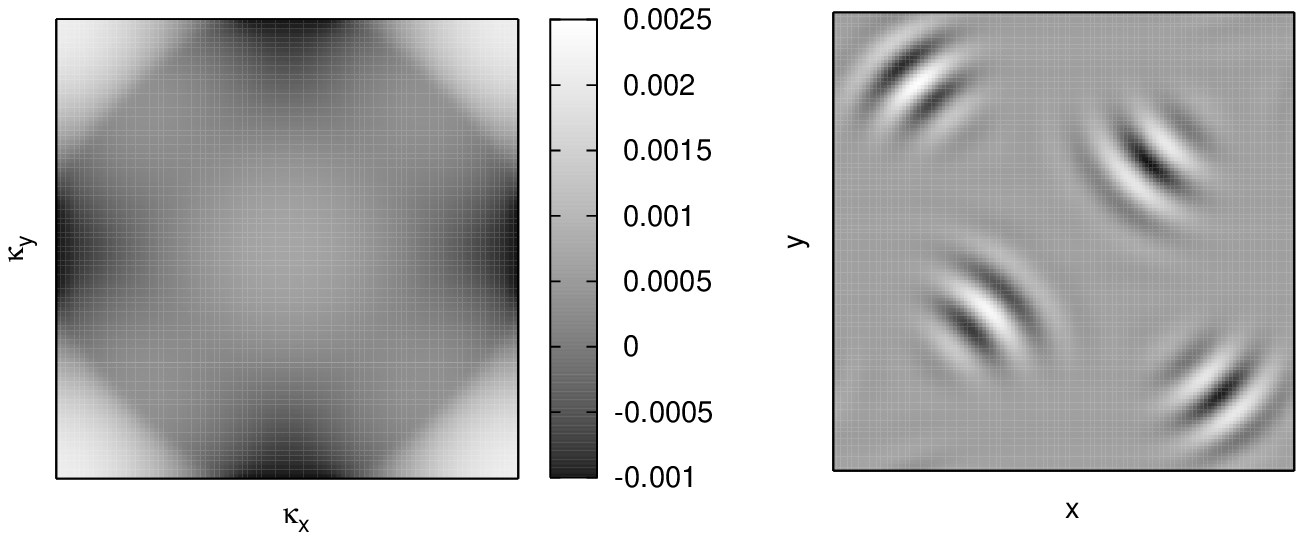}
\caption{Largest Lyapunov exponent as a function of $\kappa_x$,
$\kappa_y$ (left) and a corresponding pattern at a point of its
maximum (right) evaluated at $a=5.35$, $c=-6.35$, $U=0.18$. Maxima of LE
correspond to $\kappa_x=\pm0.5$, $\kappa_y=\pm0.5$.}
\label{fig:U0_18}
\end{figure}

We start with the parameters of Fig.~\ref{fig:LE_adv}. Near the threshold value of advection rate we present the dependence of the largest LE on the quasimomenta $\kappa_x$, $\kappa_y$, see Fig.~\ref{fig:U0_18} (left panel). This LE reaches its maximum value at
$\kappa_x=\pm 0.5$, $\kappa_y=\pm 0.5$, which indicates that the unstable patterns have a ``chess-board'' structure with respect to the periodicity of the original flow. A typical pattern of the field $\Phi$ is presented in Fig.~\ref{fig:U0_18} (right panel).
%
%
%
%

\begin{figure}[!h]
\centering
\includegraphics[width=12.0cm]{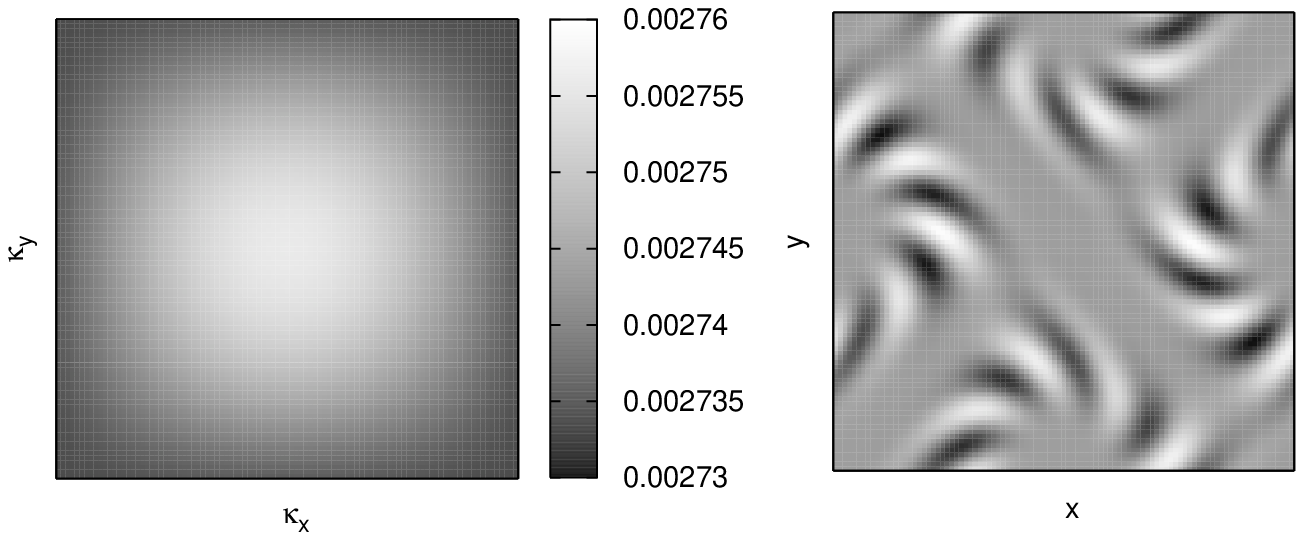}
\caption{Largest Lyapunov exponent as a function of $\kappa_x$,
$\kappa_y$ (left) and a corresponding pattern at a point of its
maximum (right) evaluated at $a=3.7$, $c=-4.7$, $U=0.77$. Maxima of LE correspond
to $\kappa_x=0$, $\kappa_y=0$.} \label{fig:U0_77}
\end{figure}

Another set of parameters is presented in Fig.~\ref{fig:U0_77}. Here, the maximum of LE corresponds to $\kappa_x=0$, $\kappa_y=0$
and the periodicity of the pattern is the same as the periodicity of the imposed flow.


\begin{figure}[!h]
\centering
\includegraphics[width=12.0cm]{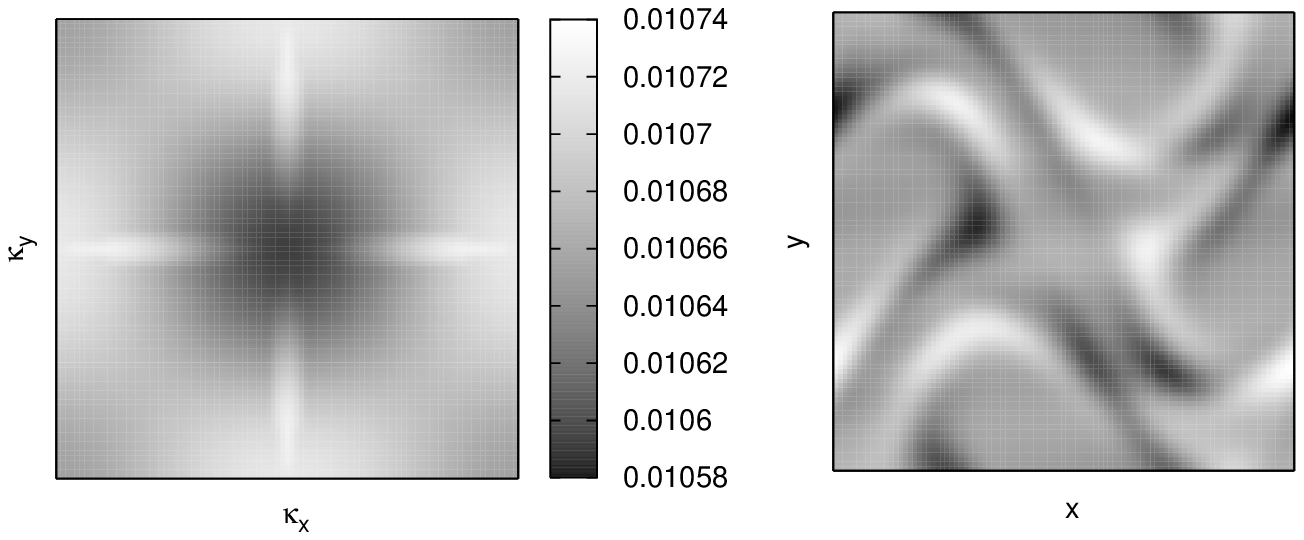}
\caption{Largest Lyapunov exponent as a function of $\kappa_x$,
$\kappa_y$ (left) and a corresponding pattern at a point of its
maximum (right) evaluated at $a=3.1$, $c=-4.1$, $U=3.5$. Maxima of LE correspond
to $\kappa_x=0$, $\kappa_y=\pm0.4$ and $\kappa_x=\pm0.4$,
$\kappa_y=0$.} \label{fig:U3_5}
\end{figure}

Finally, in  Fig.~\ref{fig:U3_5} we show the case where
the maximum of the largest LE corresponds to the points $\kappa_x=0$,
$\kappa_y=\pm0.4$ and $\kappa_x=\pm0.4$.
This is an example of a nontrivial situation where the periodicities of the patterns in $x$- and $y$- directions are not
the same and are not identical to the periodicity of the flow. We stress that in all figures above we have shown a  linear mode with the largest growth rate. On a nonlinear stage (which is beyond the scope of this paper) a pattern of the flow may significantly deviate from the linear one.


\setcounter{equation}{0}
\section{Conclusion}

The main goal of this paper is to develop an approach, based on Lyapunov-Bloch exponents, for an analysis of pattern-forming instabilities in reaction-advection-diffusion systems. It is applicable to periodic in space flows that may have arbitrary time dependence. We demonstrated the method using a discrete in time model, where reaction, advection, and diffusion act as successive operators. This enormously simplifies the calculations, while yielding a qualitatively correct picture of the process. For an exact quantitative analysis of the full system one has to apply the Lyapunov-Bloch ansatz to the full equations.

We have demonstrated that a mixing advection of one of the species may lead to a pattern-forming instability. Physically, this can be explained as crossing a threshold of Turing instability due to effective increase of one of the diffusion constants. Of course, mixing can play also a stabilizing role, suppressing spatially inhomogeneous perturbations (see, e.g., Ref.~\cite{Straube-Abel-Pikovsky-04} for such an analysis of stabilizing role of advection for chaotic in time reaction). Nonlinear patterns beyond the transition deserve further investigation, which goes beyond the scope of this paper.


\section*{Acknowledgments}
We thank M. Abel and U. Feudel for fruitful discussions. A.S. was supported by German Science Foundation, DFG SPP 1164 ``Nano- and Microfluidics,'' Project-No. STR 1021/1-2, which is gratefully acknowledged.


\appendix
\setcounter{equation}{0}

\section{Advection-diffusion map with the Bloch ansatz} \label{appendix2}

Consider the transformation of a scalar field $\phi$ due to advection, according to the map (\ref{advmap-x}), (\ref{advmap-y}). We assume that
\begin{equation}
P(x,y,t_n)=e^{i\kappa_x x+i\kappa_y y}\Phi(x,y,t_n) \nonumber
\end{equation}
\noindent and
\begin{equation}
\Phi(x,y,t_n)=\sum_{l,m}\xi_{lm}(t_n)\, e^{i(lx+my)}. \nonumber
\end{equation}
\noindent As the map is required to be area-preserving, we can write
for any iteration $\bar{P}\, d\bar{x}d\bar{y}= P\, dx dy$ (where $\bar{P}=P(\bar{x},\bar{y})$, $\bar{x}=x(t_n+T)$,
$\bar{y}=y(t_n+T)$) or
\begin{equation} \label{mass_conserv}
e^{i\kappa_x \bar{x}+i\kappa_y
\bar{y}}\,\bar{\Phi}\,d\bar{x}d\bar{y}=e^{i\kappa_x x+i\kappa_y
y}\,\Phi\,dx dy.
\end{equation}

According to (\ref{advmap-x}), during the first time interval
$T_1$ we have $\bar{x}=x+U_x T_1\sin (y+\theta_x)$, $\bar{y}=y$,
and therefore (\ref{mass_conserv}) gives us
\begin{equation}
e^{i\kappa_x U_x T_1
\sin(y+\theta_x)}\,\bar{\Phi}\,d\bar{x}d\bar{y}=\Phi\,dx dy.
\nonumber
\end{equation}
Taking this result into account, for $\tilde{\phi}_{lm}=\phi_{lm}(t_n+T_1)$ we successively obtain
\begin{eqnarray} \label{app2:adv1}
\tilde{\phi}_{lm} & = & \frac{1}{(2\pi)^2}\iint \bar{\Phi} \,
e^{-i(l\bar{x}+m\bar{y})}
\,d\bar{x}d\bar{y}=\frac{1}{(2\pi)^2}\iint \Phi \,e^{-i\kappa_x
U_x T_1 \sin(y+\theta_x)} \, e^{-i(l\bar{x}+m\bar{y})} \,dx dy \nonumber \\
& = & \frac{1}{(2\pi)^2}\iint \Phi \,e^{-i\kappa_x U_x T_1
\sin(y+\theta_x)} \, e^{-i[ lx + l U_x T_1\sin (y+\theta_x)
+m\bar{y}]} \,dx dy \nonumber \\
& = & \frac{1}{(2\pi)^2}\sum_{p,q}\phi_{pq} \iint \,e^{-i\kappa_x
U_x T_1\sin(y+\theta_x)} \, e^{-i[ lx + l U_x T_1\sin
(y+\theta_x) +my]}\,e^{i(px+qy)} \,dx dy \nonumber \\
& = & \frac{1}{(2\pi)^2}\sum_{p,q}\phi_{pq}\int\limits_0^{2\pi}
\,e^{i(p-l)x}dx \int\limits_0^{2\pi}
e^{i(q-m)y}\,e^{-iU_xT_1(l+\kappa_x)\sin(y+\theta_x)}\,dy\nonumber
\\
& = & \frac{1}{(2\pi)^2}\sum_{p,q}\phi_{pq}\, I_1 I_2 =
\frac{1}{(2\pi)^2}\sum_{p,q}\xi_{pq} \,2\pi\,\delta_{pl} \,
2\pi\,e^{-i(q-m)\theta_x} J_{q-m}[U_xT_1(m+\kappa_x)] \nonumber \\
& = & \sum_{q} e^{-i(q-m)\theta_x} J_{q-m}[U_xT_1(m+\kappa_x)]
\xi_{lq},
\end{eqnarray}
\noindent in virtue of the integrals
\begin{eqnarray}
I_1 & = & \int\limits_0^{2\pi} \,e^{i(p-l)x}dx = 2\pi
\,\delta_{pl}, \nonumber \\
I_2 & = & \int\limits_0^{2\pi}
e^{i(q-m)y}\,e^{-iU_xT_1(l+\kappa_x)\sin(y+\theta_x)}\,dy =
e^{-i(q-m)\theta_x} \, 2\pi\ J_{q-m}[U_xT_1(l+\kappa_x)],
\nonumber
\end{eqnarray}
\noindent where the Bessel function of the first kind appears
$J_m(z)=\frac{1}{2\pi}\int_{0}^{2\pi}e^{im\zeta}\,e^{-iz\sin\zeta}d\zeta$.

During the second time interval $T_2$ we have $\bar{x}=x$,
$\bar{y}=y+U_y T_2\sin (\bar{x}+\theta_y)$, which according to
(\ref{mass_conserv}) leads to an equality
\begin{equation}
e^{i\kappa_y U_y T_2
\sin(x+\theta_y)}\,\bar{\Phi}\,d\bar{x}d\bar{y}=\Phi\,dx dy.
\nonumber
\end{equation}
\noindent From an analogous consideration we obtain for
$\bar{\phi}_{lm}=\tilde{\tilde{\phi}}_{lm}=\tilde{\phi}_{lm}(t_n+T_1)=\phi_{lm}(t_n+T_1+T_2)=\phi_{lm}(t_n+T)$:
\begin{equation} \label{app2:adv2}
\bar{\phi}_{lm}=\sum_{p} e^{-i(p-l)\theta_y}
J_{p-l}[U_yT_2(m+\kappa_y)] \tilde{\phi}_{pm}.
\end{equation}

Thus, combining (\ref{app2:adv1}), (\ref{app2:adv2}) we obtain the
map describing action of advection operator $L_A$ within time
interval $T$
\begin{equation} \label{app2:fulladv}
L_A{\phi}_{lm}=\sum_{p,q} e^{-i(q-m)\theta_x} \,
e^{-i(p-l)\theta_y} J_{q-m}[U_xT_1(p+\kappa_x)]
J_{p-l}[U_yT_2(m+\kappa_y)] \phi_{pq}.
\end{equation}

Now consider the action of the diffusion operator
$L_D=\exp(TD\nabla^2)$, describing diffusion spreading of the
passive scalar with the diffusivity $D$ within time interval $T$.
Because this operator is diagonal, it acts on each Fourier
component independently:
\begin{equation} \label{app2:diff}
L_D{\phi}_{lm}=e^{-[(l+\kappa_x)^2+(m+\kappa_y)^2]DT}
\phi_{lm}.
\end{equation}

The action of both advection (\ref{app2:fulladv}) and diffusion
(\ref{app2:diff}) operators leads to the following transformation
of the Fourier amplitudes
\begin{eqnarray}
L_D L_A{\phi}_{lm} & = & e^{-[(l+\kappa_x)^2+(m+\kappa_y)^2]D
T} \sum_{p,q} g_{lmpq} \, \phi_{pq} , \label{app2:advdiff-map}\\
g_{lmpq} & = & e^{-i(q-m)\theta_x}e^{-i(p-l)\theta_y}
J_{q-m}[U_xT_1(p+\kappa_x)]J_{p-l}[U_yT_2(m+\kappa_y)]. \nonumber
\end{eqnarray}

\end{document}